\documentclass[aoas,preprint]{imsart}

\startlocaldefs
\usepackage[authoryear]{natbib}
\usepackage{graphicx}
\usepackage{color}
\usepackage{enumerate}
\usepackage{amsmath, amssymb}
\usepackage{array,multirow}
\usepackage{url}
\newtheorem{theo}{Theorem}
\newtheorem{prop}[theo]{Proposition}
\newtheorem{coro}[theo]{Corollary}
\newcommand{\bproof}{\noindent {\sl Proof.}\;}
\newcommand{\eproof}{$\square$ \\}
\renewcommand{\binom}[2]{\left(\begin{array}{c} #1 \\ #2 \end{array} \right)}
\newcommand{\thetabf}{\text{\mathversion{bold}{$\theta$}}}

\newcommand{\Ibb}{\mathbb{I}}
\newcommand{\Ybf}{{\bf Y}}
\newcommand{\mbf}{{\bf m}}
\newcommand{\E}{{E}}
\newcommand{\Eo}{{E}_0}
\newcommand{\Ea}{{E}_1}
\newcommand{\NB}{{\mathcal NB}}

\newcommand{\M}{{\mathcal M}}
\newcommand{\U}{{\mathcal U}}

\newcommand{\K}{{\mathbf K}}

\endlocaldefs

\begin{document}

\begin{frontmatter}

\title{Comparing change-point locations of independent profiles with application to gene annotation}
\runtitle{Comparing change-point locations of independent profiles}


\author{\fnms{Alice} \snm{Cleynen}\corref{}\ead[label=e1]{alice.cleynen@agroparistech.fr}},
\and
\author{\fnms{St\'ephane} \snm{Robin}\ead[label=e2]{stephane.robin@agroparistech.fr}}

\begin{aug}
\affiliation{AgroParisTech and INRA}
\address{AgroParisTech, UMR 518, and \\16 rue Claude Bernard,\\ 75005 Paris, France.\\
\printead{e1}}
\and
\address{INRA, UMR 518, \\16 rue Claude Bernard,\\ 75005 Paris, France.\\
\printead{e2}}
\end{aug}


\begin{abstract}
We are interested in the comparison of transcript boundaries from
cells which originated in different environments. The goal is to
assess whether this phenomenon, called alternative splicing, is used
to modify the transcription of the genome in response to stress
factors. We address this question by comparing the change-points
locations in the individual segmentation of each profile, which
correspond to the RNA-Seq data for a gene in one growth
condition. This requires the ability to evaluate the uncertainty of
the change-point positions, and the work of \citep{rigaill_exact_2011}
provides an appropriate framework in such case. Building on their
approach, we propose two methods for the comparison of change-points,
and illustrate our results on a dataset from the yeast specie. We show
that the UTR boundaries are subject to alternative splicing, while the
intron boundaries are conserved in all profiles. Our approach is
implemented in an R package called EBS which is available on the
CRAN. 
\end{abstract}



\end{frontmatter}

\section{Introduction}

Segmentation problems arise in a large range of domains such as economy, biology or meteorology, to name a few. Many methods have been developed and proposed in the literature in the last decades to detect change-points in the distribution of the signal along one single series. Yet, more and more applications require the analysis of several series at a time to better understand a complex underlying phenomenon. Such situations refer for example to the analysis of the genomic profiles of a cohort of patients \citep{picard2011joint}, of meteorological series observed in different locations \citep{ehsanzadeh2011simultaneous} or of sets astronomical series of photons abundance \citep{dobigeon2007joint}.

When dealing with multiple series, two approaches can be typically considered. The first consists in the {\it simultaneous} segmentation of all series, looking for changes that are common to all of them. This approach amounts to the segmentation of one single multivariate series but might permit the detection of change-points in series with too low a signal to allow their analysis independently. The second approach consists in the {\it joint} segmentation of all the series, each having its specific number and location of changes. This allows to account for dependence between the series without imposing that the changes occur simultaneously. \newline

We are interested here in a third kind of statistical problem, which is the comparison of change-point locations in several series that have been segmented separately. To our knowledge, this problem has not yet been fully addressed.

Indeed, comparing change-point is connected to the evaluation of the uncertainty of the change-point positions. An important point is that the standard likelihood-based inference is very intricate, since the required regularity conditions for the change-point parameters are not satisfied \citep{Fed75}. Most methods to obtain change-point confidence intervals are based on their limit distribution estimators \citep{Fed75, BaP03} or the asymptotic use of a likelihood-ratio statistic \citep{Mug03}. Bootstrap techniques have also been proposed (see \cite{HuK08} and references therein). 
Comparison studies of some of these methods can be found in \cite{bib_climate} for climate applications or in \cite{ToL03} for ecology.
Recently, \cite{rigaill_exact_2011} proposed a Bayesian framework to derive the posterior distributions of various quantities of interest --~including change-point locations~-- in the context of exponential family distributions with conjugate prior.

{As for the comparison of change-points, the most common approaches rely on classification comparison techniques such as the Rand Index \citep{rand1971objective}; and aim at assessing the performances of segmentation methods on single datasets, by comparing their outputs between themselves or using  the truth as reference. The notion of change-point location difference as a quantity of interest has, to our knowledge, never been considered. }

Our work is a generalization of \cite{rigaill_exact_2011} to the comparison of change point location. It is motivated by a biological problem detailed in the next paragraph.


\paragraph{Differential splicing in yeast}
Differential splicing is one of the mechanism that living cells use to modify the transcription of their genome in response to some change in their environment, such as a stress. More precisely, differential splicing refers to the ability for the cell to choose between versions (called isoforms) of a given gene by changing the boundaries of the regions to be transcribed.

New sequencing technologies, including RNA-Seq experiments, give access to a measure of the transcription at the nucleotide resolution. The signal provided by RNA-Seq consists in a count (corresponding to a number of reads) associated to each nucleotide along the genome. This count is proportional to the transcription level of the nucleotide. This technology therefore allows to locate precisely the boundaries of the transcribed regions, to possibly revise the known annotation of the genomes and to study the variation of these boundaries across conditions. 

We are interested here in an RNA-Seq experiment made on a given specie, yeast, grown under several conditions. The biological question to be addressed is 'Does yeast use differential splicing of a given gene as a response to a change in its environment?'.

\paragraph{Contribution}

In this paper we develop a Bayesian approach to compare the change-point location of independent series corresponding to the same gene {under several conditions}. We suppose that we have information on the structure of this gene (such as the number of introns) so that the number of segments of each segmentation is assumed to be known. In Section \ref{model}, we recall the Bayesian segmentation model introduced in \cite{rigaill_exact_2011} and its adaptation to our framework. In Section \ref{sec:cred} we derive the posterior distribution of the shift between the change-point locations in two independent profiles, while in Section \ref{method} we {introduce the calculation of the posterior probability for change-points to share the same location in different series. The} performances are assessed in Section \ref{simuls} via a simulation study designed to mimic real RNA-Seq data. We finally apply the proposed methodology to study the existence of differential splicing in yeast in Section \ref{appli}. Our 
approach is implemented in an R package \texttt{EBS} which is available on the CRAN repository.

All the results we provide are given conditional on the number of segments in each profiles. Indeed comparing the location of, say, the second change-points in each series implicitly refers to a total number of change-points in each of them. Yet, most of the results we provide can be marginalized over the number of segments.

\section{Model for one series} \label{model}

In this section we introduce the general Bayesian framework for the segmentation of one series and recall preceding results on the posterior distribution of change-points.

\subsection{Bayesian framework for one series}

The general segmentation problem consists in partitioning a signal
of $n$ data-points $\{y_t\}_{t \in [\![1, n]\!]}$ into $K$ segments. The model is defined as follows: the observed data
$\{y_t\}_{t=1,\ldots,n}$ are supposed to be a realization of an
independent random process $Y=\{Y_t\}_{t=1,\ldots,n}$. This process
is drawn from a probability distribution $\mathcal{G}$ which depends on a set of parameters among which one parameter $\theta$ is assumed to be affected by $K-1$ abrupt changes, called change-points and denoted $\tau_k$ ($1 \leq k \leq K-1$). A partition $m$ is defined as a set of change-points: $m = (\tau_0,\tau_1,\dots,\tau_{K})$ with conventions $\tau_0=1$ and $\tau_{K}=n+1$ and a segment $J$ is said to belong to $m$ if $J = [\![\tau_{k-1};\tau_{k}[\![$ for some $k$. \newline

The Bayesian model is fully specified with the following distributions:
\begin{itemize}
 \item the prior distribution of the number of segments $P(K)$;
 \item the conditional distribution of partition $m$ given $K$: $P(m|K)$;
 \item the parameters $\theta_J$ for each segment $J$ are supposed to be independent with same distribution $P(\theta_J)$;
 \item the observed data $Y = (Y_t)$ data are independent conditional on $m$ and $(\theta_J)$ with distribution depending on the segment: 
 $$
 (Y_t | m, J\in m, \theta_J, t\in J) \sim \mathcal{G}(\theta_J,\phi)
 $$
 where 
 $\phi$ is some parameter that is constant across the segments that will be supposed to be known. 
\end{itemize}

\subsection{Exact calculation of posterior distributions}\label{distributions}
\citet{rigaill_exact_2011} show that if distribution $\mathcal{G}$ possesses conjugate priors for $\theta_J$, and if the model satisfies the factorability assumption, that is, if 
\begin{eqnarray} \label{Eq:PYm}
P(Y,m) & = & C \prod_{J \in m} a_J P(Y_J|J), \nonumber \\
\text{where} \qquad 
P(Y_J|J) & = & \int P(Y_J|\theta_J)P(\theta_J)d\theta_J, 
\end{eqnarray}
quantities such that $P(Y,K)$, posterior change-point location distributions or the posterior entropy can be computed exactly and in a quadratic time. Examples of satisfying distributions are 
\begin{itemize}
\item the Gaussian heteroscedastic: 
$$ \mathcal{G}(\theta_J,\phi) =\mathcal{N}(\mu_J,\sigma^2_J) \; \text{with} \; \theta_J=(\mu_J, \sigma^2_J), \ \phi=\emptyset ,$$
\item the Gaussian homoscedastic with known variance $\sigma^2$: 
$$\mathcal{G}(\theta_J,\phi) =\mathcal{N}(\mu_J,\sigma^2)\; \text{with} \; \theta_J=\mu_J, \ \phi=\sigma^2,$$
\item the Poisson: 
$$\mathcal{G}(\theta_J,\phi) =\mathcal{P}(\lambda_J)\; \text{with} \; \theta_J=\lambda_J, \ \phi=\emptyset,$$ 
\item or the negative binomial homoscedastic with known dispersion $\phi$: 
$$\mathcal{G}(\theta_J,\phi) =\mathcal{NB}(p_J,\phi)\; \text{with} \; \theta_J=p_J, \ \phi=\phi.$$
\end{itemize}
Note that the Gaussian homoscedastic does not satisfy the factoriability assumption if $\sigma$ is unknown, and that the negative binomial heteroscedastic does not belong to the exponential family and does not have a conjugate prior on $\phi$. \\
The factorability assumption \eqref{Eq:PYm} also induces some constraint on the distribution of the segmentation $P(m|K)$. In this paper, we will limit ourselves to the uniform prior:
$$
P(m|K) = \mathcal U\left(\mathcal M_{K}^{1,n+1}\right)
$$
where $\M_{K}^{1,n+1}$ stands for the set of all possible partitions of $[\![1,n+1[\![$ into $K$ non-empty segments. 

\section{Posterior distribution of the shift}\label{sec:cred}

The framework described above allows to compute a set of quantities of interest in an exact manner. In this paper, we are mostly interested in the location of change-points. We first remind how posterior distributions can be computed and then propose a first exact comparison strategy.

\subsection{Posterior distribution of the change-points}

The key ingredient for most of the calculations is the $(n+1) \times (n+1)$ matrix $A$ that contains the probabilities of all segments:
\begin{eqnarray} \label{eq:matrixA}
 \forall 1\leq i < j \leq n+1,  \qquad [A]_{i,j}= P(Y_{[\![i,j[\![}|[\![i,j[\![)
\end{eqnarray}
where $P(Y_J|J)$ is given in \eqref{Eq:PYm}. 

The posterior distribution of change-points can be deduced from this matrix in a quadratic time with the following proposition:

\begin{prop} \label{Prop:PostTauk}
  Denoting $p_k(t;Y; K) = P(\tau_{k}=t|Y,K)$ the posterior distribution of the ${k}${th} change-point, we have
  $$
  p_k(t;Y;K) = \dfrac{\left[(A)^{k}\right]_{1,t}\left[(A)^{K-k}\right]_{t,n+1}}{\left[(A)^{K}\right]_{1,n+1}}.
  $$
\end{prop}

\bproof 
We have $$ p_k(t;Y;K) =\frac{\sum_{m\in \mathcal{B}_{K,k}(t)}p(Y|m)p(m|K)}{P(Y|K)}$$ where $\mathcal{B}_{K,k}(t)$ is the set of partitions of $\{1,\dots,n\}$ in $K$ segments with $k$th change-point at location $t$. 
{Note that $\mathcal{B}_{K,k}(t)=\mathcal{M}_k^{1,t}\otimes\mathcal{M}_{K-k}^{t,n+1}$ (i.e. all $m \in \mathcal{B}_{K,k}(t) $ can be decomposed uniquely as $m=m_1\cup m_2$ with $m_1 \in \mathcal{M}_k^{1,t}$ and $m_2 \in \mathcal{M}_{K-k}^{t,n+1}$ and reciprocally).}
Then using the factoriability assumption, we can write 
$$ p_k(t;Y;K) =\dfrac{\sum_{m_1\in \mathcal{M}_k^{1,t}}p(Y|m_1)\sum_{m_2\in \mathcal{M}_{K-k}^{t,n+1}}p(Y|m_2)\; p(m|K)}{\sum_{m \in \M_{K}^{1,n+1}}p(Y|m)\; p(m|K)}$$
\eproof

\subsection{Comparison of two series}

We now propose a first procedure to compare the location of two change-points in two independent series. Consider two independent series $Y^1$ and $Y^2$ with same length $n$ and respective number of segments $K^1$ and $K^2$. The aim is to compare the locations of the $k_1$th change-point from of series $Y^1$ (denoted $\tau_{k_{1}}^{1}$) with the $k_2$th change-point of  series $Y^2$ (denoted $\tau_{k_{2}}^{2}$). The posterior distribution of the difference between the location of the two change-points can be derived with the following Proposition.

\begin{prop} \label{Prop:PostDelta}
  Denoting $\delta_{k_1, k_2}(d; K^1, K^2) = P(\Delta=d|Y^1, Y^2, K^1, K^2)$ the posterior distribution of the difference $\Delta =  \tau_{k_{1}}^{1}-\tau_{k_{2}}^2$, we have
  $$
  \delta_{k_1, k_2}(d; K^1, K^2) =\sum_t p_{k_1}(t;Y^1; K^1) p_{k_2}(t-d ;Y^2; K^2).
  $$
\end{prop}

\bproof
This simply results from the convolution between the two posterior distributions $p_{k_1}$ and $ p_{k_2}$. 
\eproof

The posterior distribution of the shift can therefore be computed exactly and in a quadratic time. The non-difference between the two change-point locations $\tau_{k_{1}}^{1}$ and $\tau_{k_{2}}^{2}$ can then be assessed, looking at the position of 0 with respect to the posterior distribution $\delta$.

\section{Comparison of change point locations} \label{method}

We now consider the comparison of change-point locations between more than 2 series. In this case, the convolution methods described above does not apply anymore so we propose a comparison based on the exact computation of the posterior probability for the change-points under study to have the same location.

\subsection{Model for $I$ series}
We now consider $I$ independent series $Y^\ell$ (with $1\leq \ell \leq I$) with same length $n$. We denote $m^\ell$, their respective partitions and $K^\ell$ their respective number of segments. We further denote $\tau_k^\ell$ the $k${th} change-point in $Y^\ell$ so $m^\ell = (\tau_0^\ell, \tau_1^\ell, \dots, \tau_{K^\ell}^\ell)$. Similarly, $\theta_J^\ell$ denotes the parameter for the series $\ell$ within segment $J$ provided that $J \in m^\ell$ and $\phi^\ell$ the constant parameter of series $\ell$. In the following, the set of profiles will be referred to as $\Ybf$ and respectively for the vector of segment numbers ($\K$), the set of all partitions ($\mbf$) and the set of all parameters ($\thetabf$). 

\begin{figure}[h!]
  \begin{center}
  \includegraphics[width=.2\textwidth]{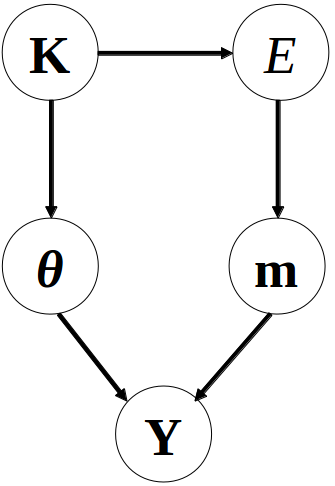}
  \caption{{\bf Graphical model.} Hierarchical model for the comparison of $I$ series.} \label{Fig:GraphModel}
  \end{center}
\end{figure}

In the perspective of change-point comparison, we introduce the following event:
$$
\Eo = \{\tau_{k_1}^{1}=\dots=\tau_{k_I}^{I}\}.
$$
We further denote $\Ea$ its complementary and define the binary random variable 
$$
\E = \Ibb\{\Ea\} = 1 - \Ibb\{\Eo\}.
$$
The complete hierarchical model is displayed in Figure \ref{Fig:GraphModel} and is defined as follows:
\begin{itemize}
 \item The random variable $\E$ is drawn conditionally on $\K$ as a Bernoulli $\mathcal B(1- p_0(\K))$ where $p_0(\K) = P(\Eo | \K)$;
 \item The parameters $\thetabf$ are drawn independently according to $P(\thetabf | \K)$;
 \item The partitions are drawn conditionally on $\E$ according to $P(\mbf | \K, \E)$;
 \item The observations are generated according to the conditional distribution $P(\Ybf | \mbf, \thetabf)$.
\end{itemize}
More specifically, denoting $\M_{\K}^{1,n+1} = \bigotimes_\ell \M_{K^\ell}^{1,n+1}$, the partitions are assumed to be uniformly distributed, conditional on $E$, that is
$$
P(\mbf | \K, \Eo) = \U(\M_{\K}^{1,n+1} \cap \Eo), 
\qquad
P(\mbf | \K, \Ea) = \U(\M_{\K}^{1,n+1} \cap \Ea).
$$

\subsection{Posterior probability for the existence of a common change-point}

We propose to assess the existence of a common change-point location between the $I$ profiles based on the posterior probability of this event, namely $P(\Eo | \Ybf, \K).$

\begin{prop} \label{Prop:PosteriorE0}
  The posterior probability of $\Eo$ can be computed in $O(Kn^2)$ as
  \begin{eqnarray*} 
  P(\Eo | \Ybf, \K) & = & \frac{p_0(\K)}{q_0(\K)} Q(\Ybf, \Eo | \K)\; . \\
  && \left[ \frac{1 - p_0(\K)}{1 - q_0(\K)} Q(\Ybf | \K) + \frac{p_0(\K) - q_0(\K)}{q_0(\K)[1 - q_0(\K)]} Q(\Ybf, \Eo | \K) \right]^{-1} 
  \end{eqnarray*}
  where
  \begin{eqnarray*} 
  Q(\Ybf | \K) & = & \prod_\ell \left[(A_\ell)^{K_\ell}\right]_{1, n+1}, \\
   Q(\Ybf, \Eo | \K) & = & \sum_t \prod_\ell  \left[(A_\ell)^{k_\ell}\right]_{1, t} \left[(A_\ell)^{K_\ell-k_\ell}\right]_{t+1, n+1},\\
  \text{and } q_0(\K)  =  Q(\Eo|\K) &=& \sum_t \prod_\ell \begin{small} \binom{t-2}{k_\ell-1} \binom{n-t}{K_\ell-k_\ell-1} \left/ \binom{n-1}{K_\ell-1} \right. \end{small} .
  \end{eqnarray*}
  and $A_\ell$ stands for the matrix $A$ as defined in (\ref{eq:matrixA}), corresponding to series $\ell$.
\end{prop}

\bproof
We consider the surrogate  model where the partition $\mbf$ is drawn uniformly and independently from $\E$, namely $Q(\mbf | \K) = \U(\M_{\K}^{1,n+1})$ (note that this corresponds to choosing
$p_0(\K) = q_0(\K)$). All probability distributions under this model are denoted by $Q$ along the proof. 
The formulas for probabilities $Q(\Ybf | \K)$ and $Q(\Ybf, \Eo | \K)$ derive from \cite{rigaill_exact_2011}. It then suffices to apply the probability change as
$$
P(\Ybf, \Eo | \K) = \frac{p_0(\K)}{q_0(\K)} Q(\Ybf, \Eo | \K), 
\quad 
P(\Ybf, \Ea |\K) = \frac{1 - p_0(\K)}{1 - q_0(\K)} Q(\Ybf, \Ea |\K).
$$
The result then follows from the decomposition of $P(\Ybf | \K)$ as $P(\Ybf, \Eo |\K) + P(\Ybf, \Ea |\K)$  and the same for $Q(\Ybf | \K)$.
\eproof

The Bayes factor is sometimes preferred for model comparison; it can be computed exactly in a similar way:

\begin{coro}
  The Bayes factor can be computed in $O(Kn^2)$ as
  \begin{eqnarray*} 	
  \frac{P(\Ybf | \Eo, \K)}{P(\Ybf |\Ea, \K)} & = & \frac{1 - q_0(\K)}{q_0(\K)} \; \frac{Q(\Ybf, \Eo | \K)}{Q(\Ybf | \K) - Q(\Ybf, \Eo | \K)}
  \end{eqnarray*}
  using the same notations as in Proposition \ref{Prop:PosteriorE0}.
\end{coro}

\bproof
The proof follows this of Proposition \ref{Prop:PosteriorE0}.
\eproof

\section{Simulation study}\label{simuls}

\subsection{Simulation design}\label{description}
\begin{figure}[h!]
\center{\includegraphics[width=10cm]{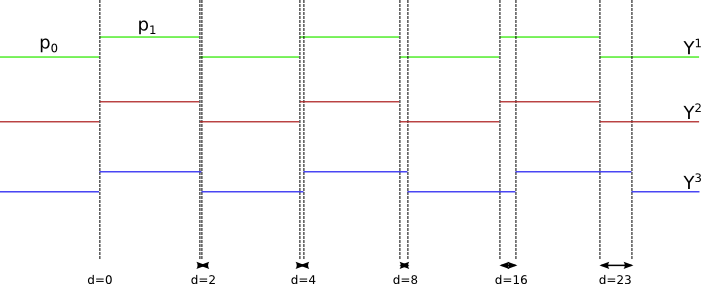}}
\caption{{\bf Simulation design.}}\label{plansimu}
\end{figure}

We designed a simulation study to identify the influence of various parameters on the performances of our approach.  The design is illustrated in Figure \ref{plansimu}: we compared $3$ independent profiles with $7$ segments, with all odd (respectively even) segments sharing the same distribution. The first two profiles have identical segmentation $m$ given by $m=(1,101,201,301,401,501,601,701)$ and the change-point locations of the third one are progressively shifted apart as $\tau_{k}^3=\tau_{k}^1+2^{k-1}$, for each $1\leq k \leq 6$. We shall denote $d_k=\tau_k^3-\tau_k^1$ and drop the index $k$ when there is no ambiguity on it. \newline

Our purpose is to mimic data obtained by RNA-Seq experiments, so that the parameters for the negative binomial distribution were chosen to fit typical real-data.  Considering the model where odd segments are sampled with distribution $\NB(p_0,\phi)$, and even with $\NB(p_1,\phi)$, we chose two different values of $p_0$, $0.8$ and $0.5$, and for each of them, we made $p_1$ vary so that the odd-ratio $s := p_1 /(1-p_1) / [p_0 / (1-p_0)]$ is $4$, $8$ and $16$. Finally, we used different values of $\phi$ as detailed in Table \ref{paramvalue} in order to explore a wide range of possible dispersions while keeping a signal/noise ratio not too high. Note that the higher $\phi$, the less overdispersed the signal. From our experience, the configuration of parameter combinations with $p_0=0.5$ is the more typical of observed values for RNA-Seq data. \newline

\begin{table}[h!]
\begin{center}
\begin{tabular}{|c|c|c|c|}
\multicolumn{2}{|c|}{$p_0=0.8$} & \multicolumn{2}{c|}{$p_0=0.5$}\\
\hline
 $p_1$ & $\phi$ & $p_1$ & $\phi$ \\
 \hline
 $0.5$ & $5$ & $0.2$ & $0.08^{1/8}$\\
 $0.33$ & $\sqrt{5}$ & $0.1$ & $0.08^{1/4}$ \\
 $0.2$ & $0.8$ & $0.05$ & $0.08^{1/2}$ \\
  & $0.64$ & & $0.08$\\
\end{tabular}
\caption{{\bf Values of parameters used in the simulation study}}\label{paramvalue}
\end{center}
\end{table}

{Provided that the ratio $\lambda=\phi(1-p)/p$ remains constant, the negative binomial distribution with dispersion parameter $\phi$ going to infinity converges to the Poisson distribution $\mathcal{P}(\lambda)$. We propose an identical simulation study based on the Poisson distribution for the comparison with non-dispersed datasets. Specifically, we used for $\lambda_0$ the values $1.25$ and $0.73$ so that the odd-ratios $s=4;8;16$ corresponded to the respective values $\lambda_1=5;10;20$ and $2.92; 5.83;11.7$} \newline

In practice there is little chance that the overdispersion is known. We propose to estimate this parameter from the data and use the obtained value in the analysis. The results presented here used the estimator inspired from \cite{jonhson_kotz}: starting from sliding window of size $15$, we compute the method of moments estimator of $\phi$, using the formula $\phi = \E^2(X)/(V(X)-\E(X))$, and retain the median over all windows. When this median is negative (which is likely to happen in datasets with many zeros), we double the size of the window. In practice however, results are very similar when using maximum likelihood or quasi-maximum likelihood estimators on sliding windows. 
 \newline

\subsection{Results}

We compute the posterior probability $P(\Eo | \Ybf, \K)$ for each simulation and each value of $d$. Figures \ref{Pois} to \ref{BN2} in Appendix \ref{abacus} represent the boxplots of this probability for each configuration. For sake of visibility, the outliers were not drawn in those figures. Note that in each figure, the first boxplot corresponds to $d=0$ and thus to model $\Eo$, while $d\neq 0$ for left boxplots so that the true model is $\Ea$. These plots can be understood as abacus for the detection power of the proposed approach. For example, the perfect scenario corresponds to $s=16$ in the Poisson case of Figure \ref{Pois}.   \newline

As expected, these results show that the lower the value of $\phi$ (the Poisson distribution is interpreted here as  $\phi = +\infty$), the most difficult the decision becomes. The trend is identical for decreasing values of the odd-ratio $s$ and decreasing values of $d$. 
In the most difficult scenario of very high dispersion compared to signal value, the method fails to provide satisfying decisions whatever the level of odd-ratio or distance between change-points. However, in most configurations, the method is adequate as soon as $d\geq 16$. \newline

\begin{figure}[h]
 \centerline{\includegraphics[width=12cm]{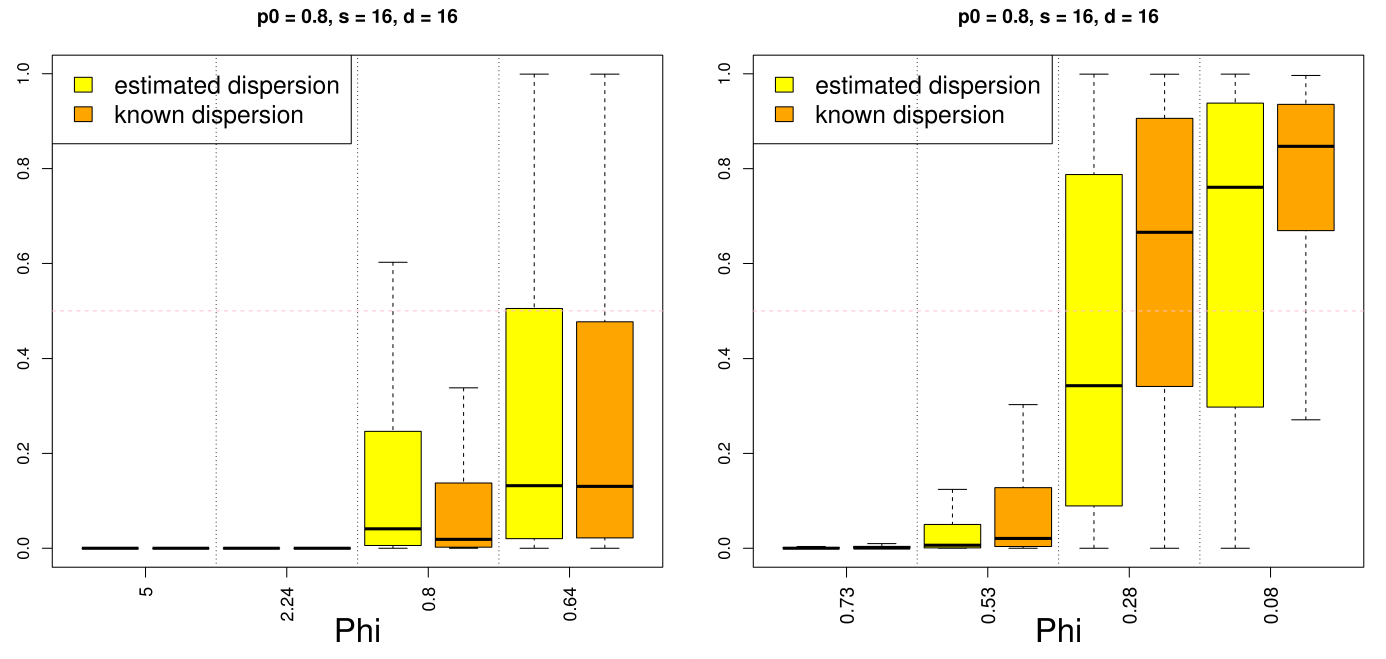} }
 \caption{{\bf Impact of estimating the dispersion parameter.} Boxplot of the posterior probability of $\Eo$ for $s=16$ and $d=16$ when estimating the value of $\phi$ (left boxplot of each subdivision) or when using the known value (right boxplot of each subdivision).}\label{phi}
\end{figure}

An important question is the impact of the estimation of the dispersion parameter. Interestingly, in the simulation study with $p_0=0.8$, our estimator tended to under-estimate $\phi$ (and thus over-estimate the dispersion) while it was the contrary in the simulation study with $p_0=0.5$. This affects the performance of the decision rule, which behaves better when $\phi$ is higher. For instance, Figure \ref{phi} shows, for $s=16$ and $d=16$, that knowing the true value of $\phi$ improves the results when $p_0=0.8$ but worsens them when $p_0=0.5$.

\section{Comparison of transcribed regions in yeast} \label{appli}

\paragraph{Experimental design.}
We now go back to our first motivation a consider a study from the Sherlock lab in Stanford \citep{Risso_norma}. In their experiment, they grew a yeast strain, \textit{Saccharomyce Cerevisiae}, in three different environments: ypd, which is the traditional (rich) media for yeast, delft, a similar but poorer media, and glycerol. {In the last decade many studies (see for instance \cite{polyA1,Tian-polyA}) have showed that a large proportion of genes have more than one polyadenylation sites, thus can express multiple transcripts with different 3' UTR sizes. Similarly, the 5' capping process is dependent on environment conditions  \citep{5cap}, and the 5' UTR size may vary according to stress factors. We may therefore expect that the yeast cells grown in different conditions (they ferment in the first two media, while they respire in glycerol) will produce transcripts of unequal sizes. On the contrary, the intron-exon boundaries are not expected to differ between conditions}

\begin{figure}[h]
 \centerline{\includegraphics[width=12cm]{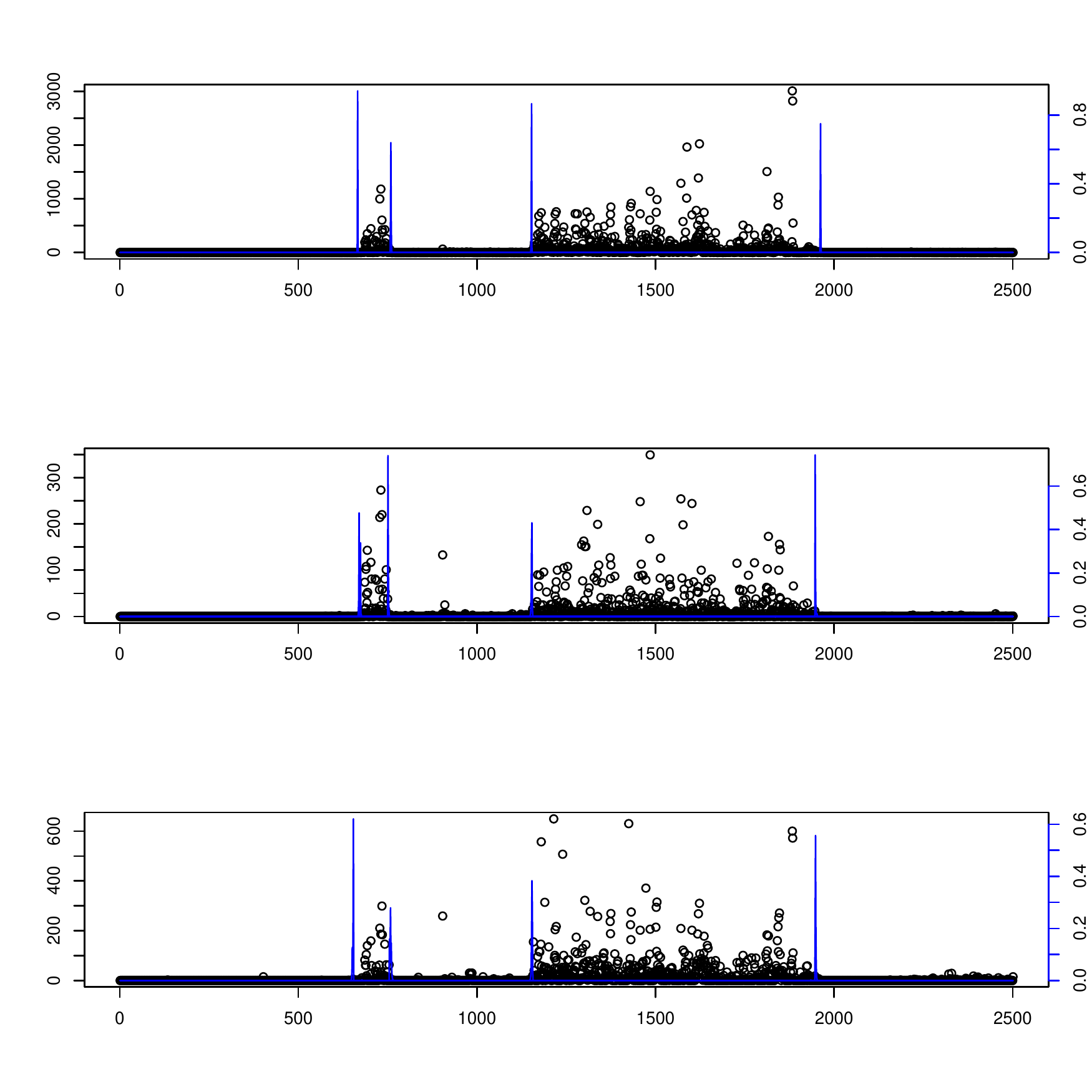}}
 \caption{{\bf Posterior distribution of change-point location.} Segmentation in $5$segments of gene YAL013W in three different media: ypd (top), delft (middle) and glycerol (bottom). Black dots represent the number of reads starting at each position of the genome (left scale) while blue curves are the posterior distribution of the change-point location (right scale).}\label{real-data}
\end{figure}

\paragraph{Change-point location.}
We applied our procedure to gene YAL013W which has two exons. The RNA-Seq series were segmented into $5$ segments to allow one segment per transcribed region separated by segments of non-coding regions. Figure \ref{real-data} illustrates the posterior distribution of each change-point in each profile.

\paragraph{Credibility intervals on the shift.}
For each of the first to the fourth change-point, we {computed the posterior distribution of the difference between change-point locations for each pairs of conditions}. {For the biological reasons stated above, we expect to observe more differences for the first and last change-points than for the other two, which can be used as a verification of the decision rule.}

\begin{figure}[h]
 \centerline{\includegraphics[width=13cm]{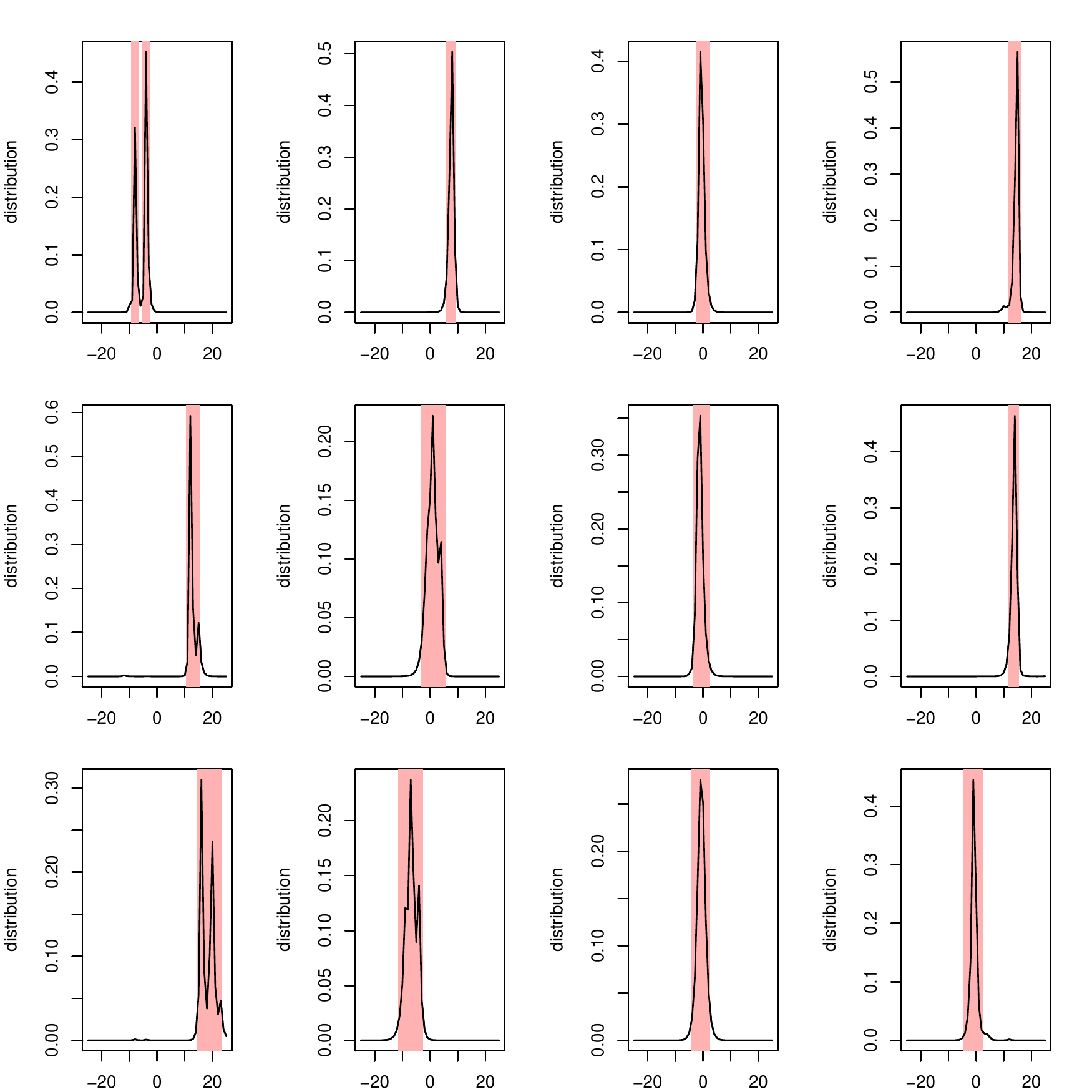}}
 \caption{{\bf Distribution of change-point location and $95$\% credibility intervals. } For each of the two by two comparison (top: ypd-delft; middle: ypd-glycerol; bottom delft-glycerol), posterior distribution of the change-point difference for each of the first to the fourth change-point.}\label{credibility}
\end{figure}

Figure \ref{credibility} {provides} the {posterior} distribution of {these differences}, as well as the $95$\% credibility intervals.

\paragraph{Posterior probability of common change-point.}
{We then computed the probability that the change-point is the same across several series, taking $p_0 = 1/2$.} Table \ref{res:BF} provides, for the simultaneous comparison of the three conditions and for each pair of conditions, the value of the posterior probability of $\Eo$ at each change-point ($\tau_1^\ell$ is associated with the $5'$ UTR, $\tau_2^\ell$ to the 5' intron boundary, $\tau_3^\ell$ to the $3'$ intron boundary and $\tau_4^\ell$ to the $3'$ UTR).  Reassuringly, in most cases the change-point location is identical when corresponding to intron boundaries. {On the contrary, UTR boundaries seem to differ from one condition to another.}

\begin{table}[h!]
\begin{center}
\begin{tabular}{ccccc}
\multirow{2}{*}{comparison} & \multicolumn{4}{c}{change-point} \\
 & $\qquad \tau_1 \qquad $ & $\qquad \tau_2 \qquad$ & $\qquad \tau_3 \qquad$ & $\qquad \tau_4 \qquad$ \\
\hline
all media & $10^{-3}$ & $0.99$ & $0.99$ & $6\;10^{-3}$ \\
ypd-delft & $0.32$ & $0.30$ &$0.99$ & $10^{-5}$ \\
ypd-glycerol & $4\;10^{-4}$ & $0.99$ &$0.99$ & $6\;10^{-3}$ \\
delft-glycerol & $5\;10^{-2}$ & $0.60$ & $0.99$ & $0.99$ \\
\hline
\end{tabular}
\caption{{Posterior probability of a common change point across conditions for gene YAL013W}}\label{res:BF}
\end{center}
\end{table}

\begin{figure}[h]
 \centerline{\includegraphics[width=13cm]{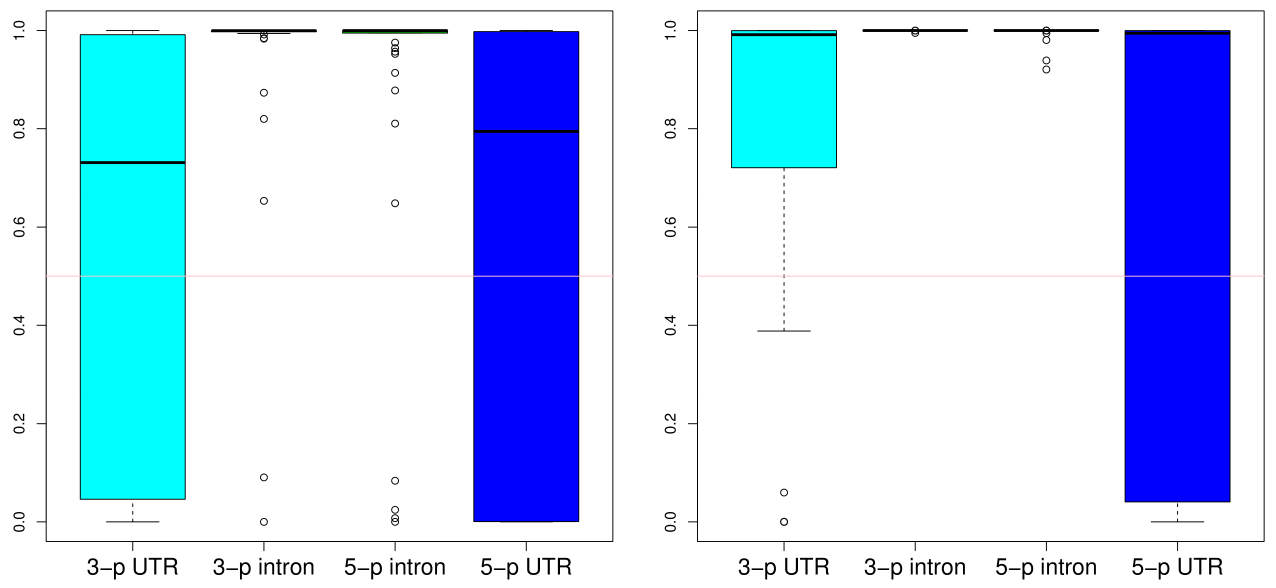} }
 \caption{{\bf Distribution of $P(\Eo | \Ybf, \K)$ for a set of $50$ genes with two values of $p_0$.}  We set $p_0=1/2$ in the left figure, and $p_0=0.9$ for $\tau_1$ and $\tau_4$, $p_0=0.99$ for the intron boundaries in the right figure.}\label{50genes}
\end{figure}

\paragraph{Differential splicing in yeast.} We finally applied our comparison procedure to a set of $50$ genes from the yeast genome which all possess two exons and which were expressed in all three conditions at the time of the experiment. The left figure of Figure \ref{50genes} shows the distribution of the posterior probability of $\Eo$  for the simultaneous comparison of the three conditions when $p_0(\K)=1/2$. Once again the results strengthens the expectation that intron boundaries should not vary between conditions while more difference is observed for the UTRs. A closer look at the five genes for which we have evidence of either the second or third change-point difference reveals that one of the two exons was not expressed in the Glycerol medium. Moreover, a discussion with Dr Sherlock suggests that about $10$\% of the genes should 	be liable to differential splicing. We therefore performed the analysis over again removing the $5$ outliers and setting $p_0=0.9$ for $\tau_1$ and $\tau_4$ and $p_0=0.99$ 
for the other two. Results are illustrated in the right figure of Figure \ref{50genes}. For these new prior values, we observe that $9$ genes have a $3'$ UTR length which varies, and $16$ for the $5'$ UTR.

\section{Conclusion}

We have proposed two exact approaches  for the comparison of change-point location. The first is based  on the posterior distribution  of the shift in  two profiles, while the second is adapted to the comparison of multiple profiles and studies the posterior probability of having a common change-point. These procedures, when applied to RNA-Seq datasets, confirm the expectation that transcription starting and ending sites may vary between growth conditions while the localization of introns remains the same.

While we have illustrated these procedures with count datasets, they can be adapted to all distributions from the exponential family verifying the factoriability assumption as described in Section \ref{distributions}. They are in fact implemented in an R package \texttt{EBS} for the negative binomial, Poisson, Gaussian heteroscedastic and Gaussian homoscedastic with known variance parameter. This package is available on the CRAN repository at \url{http://cran.r-project.org/web/packages/EBS/index.html}.

\section*{Acknowledgments}

The authors deeply thank Sandrine Dudoit, Marie-Pierre Etienne, Emilie Lebarbier Eric Parent and Gavin Sherlock for helpful conversations and comments on this works.

\bibliographystyle{plainnat}
\bibliography{Biblio}

\appendix
\section{Appendix section}\label{abacus}
\begin{figure}
\centerline{\includegraphics[width=12cm]{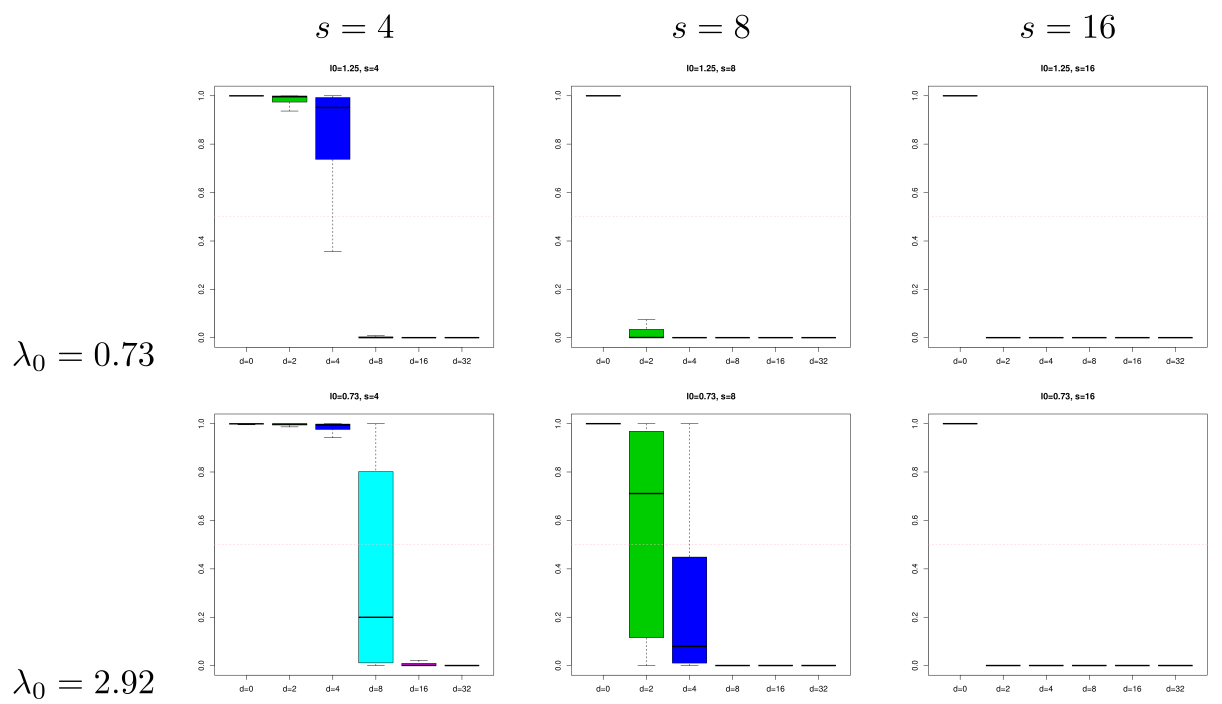} }
  \caption{{\bf Boxplot of posterior probabilities of $\Eo$ for Poisson.} Plotted as $d$ increases in simulation studies for the Poisson distribution with $\lambda_0=0.73$ (Top) and $\lambda_0=2.92$ (Bottom) and for each value of $s$ (in columns).} \label{Pois}
\end{figure}

\begin{figure}
 \centerline{\includegraphics[width=12cm]{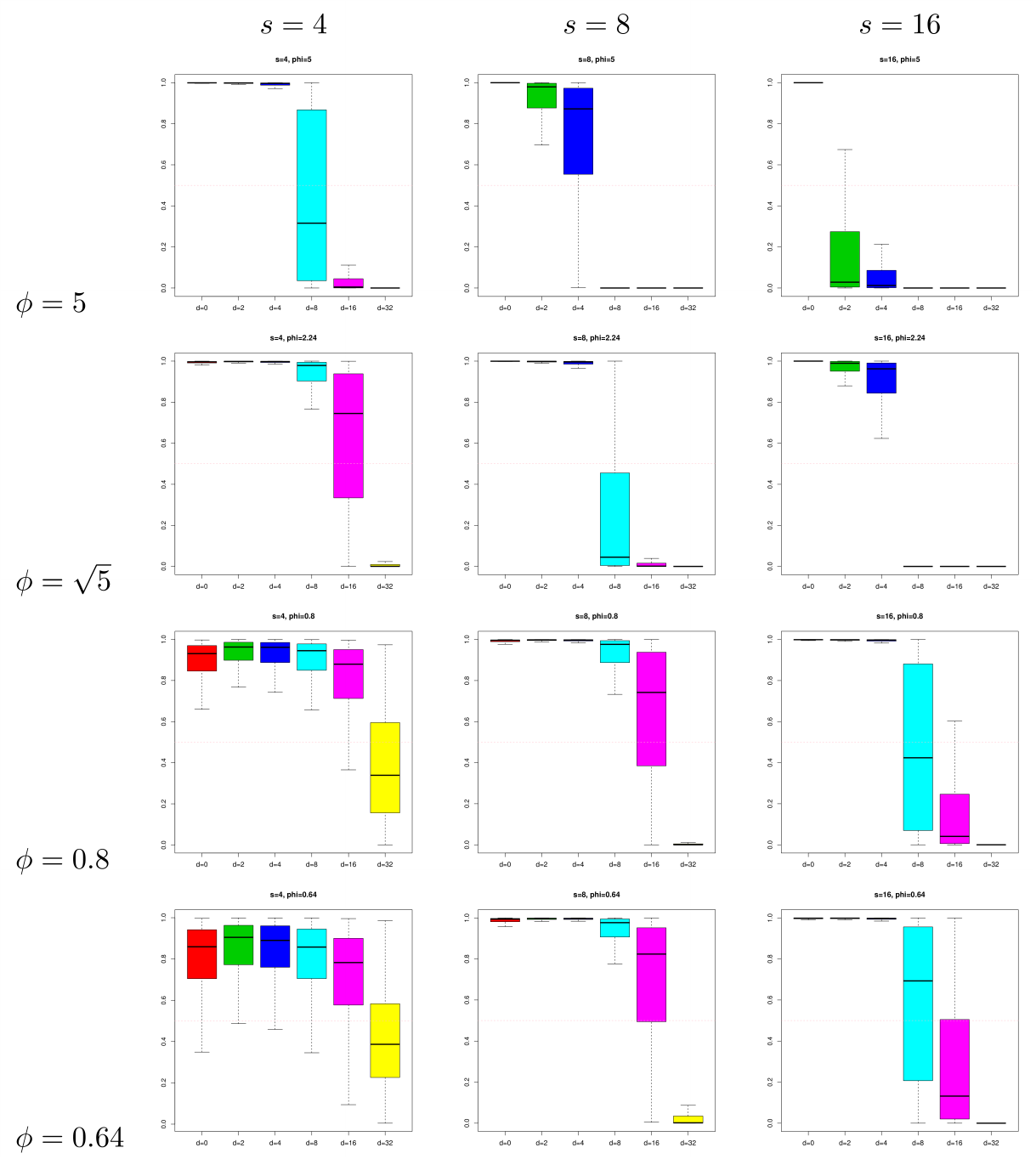} }
  \caption{{\bf Boxplot of posterior probabilities of $\Eo$ for negative Binomial, with $p_0=0.8$.} Plotted as $d$ increases in simulation studies for the negative binomial distribution  with $p_0=0.8$ and for each value of $s$ (in columns) and each value of $\phi$ (in rows) as detailed in the left side of Table \ref{paramvalue}. The overdispersion is estimated as detailed in Section \ref{description}.} \label{BN1}
\end{figure}

\begin{figure}
 \centerline{\includegraphics[width=12cm]{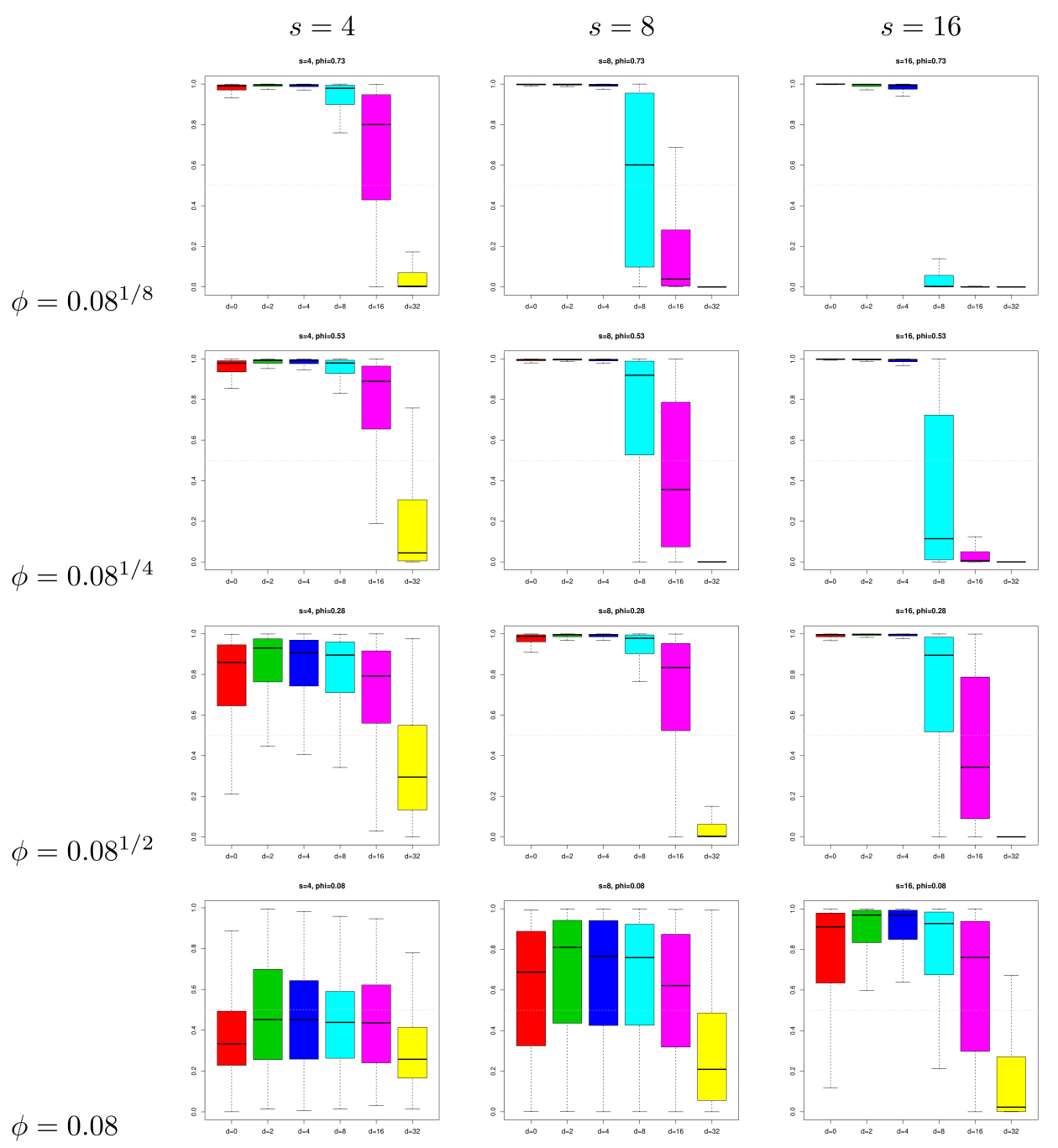} }
  \caption{{\bf Boxplot of posterior probabilities of $\Eo$ for negative Binomial, with $p_0=0.5$.} Plotted as $d$ increases in simulation studies for the negative binomial distribution with $p_0=0.5$ and for each value of $s$ (in columns) and each value of $\phi$ (in rows) as detailed in the right side of Table \ref{paramvalue}. The overdispersion is estimated as detailed in Section \ref{description}.} \label{BN2}
\end{figure}




\end{document}